\documentclass[12pt]{article}
\usepackage{authblk}
\usepackage{siunitx}
\usepackage{graphicx}
\usepackage{caption}
\usepackage{amsmath}
\usepackage{bpchem}
\usepackage{float}
\usepackage[margin=0.75in]{geometry}

\DeclareSIUnit[number-unit-product = {}]
\uL{\micro\liter}
\DeclareSIUnit[number-unit-product = {}]
\umol{\micro \mole}
\DeclareSIUnit[number-unit-product = {}]
\uM{\micro M}
\newcommand{\textsub}[1]{$_{\text{#1}}$}

\title{Microparticle assembly pathways on lipid membranes}
\date{March 31, 2017}

\author[1]{Casper van der Wel}
\author[1, 2]{Doris Heinrich}
\author[1, *]{Daniela J. Kraft}

\affil[1]{Biological and Soft Matter Physics, Huygens-Kamerlingh Onnes Laboratory, Leiden University, PO Box 9504, 2300 RA Leiden, The Netherlands}
\affil[2]{Fraunhofer Institute for Silicate Research, Neunerplatz 2, 97082 W\"urzburg, Germany}
\affil[*]{To whom correspondence should be addressed. E-mail: kraft@physics.leidenuniv.nl}

\begin{document}
\maketitle

\begin{abstract}
Understanding interactions between microparticles and lipid membranes is of increasing importance, especially for unraveling the influence of microplastics on our health and environment.
Here, we study how a short-ranged adhesive force between microparticles and model lipid membranes causes membrane-mediated particle assembly.
Using confocal microscopy, we observe the initial particle attachment to the membrane, then particle wrapping, and in rare cases spontaneous membrane tubulation.
In the attached state, we measure that the particle mobility decreases by 26 \%.
If multiple particles adhere to the same vesicle, their initial single-particle state determines their interactions and subsequent assembly pathways:
1) attached particles only aggregate when small adhesive vesicles are present in solution, 2) wrapped particles reversibly attract one another by membrane deformation, and 3) a combination of wrapped and attached particles form membrane-mediated dimers, which further assemble into a variety of complex structures. 
The experimental observation of distinct assembly pathways induced only by a short ranged membrane-particle adhesion, shows that a cellular cytoskeleton or other active components are not required for microparticle aggregation.
We suggest that this membrane-mediated microparticle aggregation is a reason behind reported long retention times of polymer microparticles in organisms.
\end{abstract}

\section*{Introduction}
Artificial microparticles are increasingly applied in ceramics, paints, cosmetics, drug delivery \cite{Tang2012}, and microbiological techniques such as microrheology \cite{Tseng2002}.
However, the negative environmental effects of polymer-based microparticles, for example through uptake by and high retention in marine organisms, are becoming increasingly clear \cite{Cole2011a}.
In order to be able to unravel the implications of microparticles on health and environment, it is crucial to understand their impact on cellular processes and especially their interactions with the cellular membrane, which is the most important protective barrier of the cell. 

The first step in the interaction of microparticles with living cells is their adhesion to the membrane.
This adhesion can be caused by a variety of mechanisms such as Van der Waals, Coulomb or hydrophobic forces, or complementary protein interactions \cite{Nel2009, Mahmoudi2014, Vonarbourg2006a}. 
Subsequent internalization into the cell depends on the cell type, the particle size, and the particle surface moieties \cite{Vonarbourg2006a, Rejman2004, Lorenz2006, Gaumet2009}.
Especially particles of micrometer size have been observed to not be internalized, but to form aggregates that remain irreversibly attached to the cellular membrane \cite{Lorenz2006, Gaumet2009}.
Intriguingly, this membrane-mediated particle aggregation has also been reported on lipid vesicles \cite{Ramos1999,Koltover1999,Sarfati2016a,VanderWel2016b}, suggesting that a physical property of the lipid membrane drives the observed particle aggregation on living cells.

Here, we investigate this membrane-mediated particle assembly using a well-controlled model system that we have developed recently \cite{VanderWel2016b}.
We study the membrane-microparticle system with confocal microscopy, which enables us to visualize the particle-induced membrane deformation, quantify the particle mobilities, and categorize subsequent assembly pathways.
Previously, we established that a single membrane-adhesive particle is either completely wrapped by the membrane, or only attaches without inducing visible deformation.
Here, we first quantify the mobilities of these two types of membrane-associated particles and find that the membrane significantly reduces the particle diffusivity.
Then we study the aggregation pathways that lead to large membrane-mediated aggregates.
We establish that the initial state of individual particles determines which assembly pathway they take: wrapped particles only interact via a deformation-mediated force, attached particles can stick together via small adhesive vesicles, and wrapped and attached particles form dimers driven by the strong membrane-particle adhesion.
Finally, we describe and discuss the case in which a wrapped particle induces spontaneous membrane tubulation.

\section*{Materials and Methods}
\subsection*{Chemicals}
Styrene (99\%), itaconic acid (99\%), 4,4'-Azobis(4-cyanovaleric acid) (98\%, ACVA),
D-glucose (99\%),
methanol (99.9\%),
chloroform (99\%),
sodium phosphate (99\%),
deuterium oxide  (70\%),
\IUPAC{N-hydroxy\|sulfo\|succinimide sodium salt} (98\%, Sulfo-NHS), and
\IUPAC{1,3,5,7-tetra\|methyl-8-phenyl-4,4-difluoro\-bora\-diaza\-indacene} (97\%, BODIPY)
were acquired from Sigma-Aldrich; 
\IUPAC{methoxy\|poly\|(ethy\|lene) glycol amine} (mPEG, M\textsub{w} = 5000) from Alfa Aesar;
sodium chloride (99\%), and
sodium azide (99\%)
from Acros Organics;
\IUPAC{1-ethyl\-3-(3-dimethyl\|amino\|propyl) carbodiimide hydrochloride} (99\%, EDC) from Carl Roth;
NeutrAvidin (avidin) and cholera toxin (subunit B) Alexa Fluor 488 conjugate
(cholera toxin 488) from Thermo Scientific;
DNA oligonucleotides (biotin-5'-TTTAATATTA-3'-Cy3) from Integrated DNA Technologies;
\IUPAC{$\Delta$9-cis 1,2-dioleoyl-\textit{sn}-glycero-3-phosphocholine} (DOPC),
\IUPAC{1,2\-dioleoyl\-\textit{sn}\-glycero\-3-phospho\|ethanol\|amine\-N\-[biotinyl\|(poly\|ethylene glycol)\-2000]} (DOPE-PEG-biotin),
\IUPAC{1,2\-dioleoyl\-\textit{sn}\-glycero\-3-phospho\|ethanol\|amine\-N-(lissamine\|rhodamine B sulfonyl)} (\IUPAC{DOPE\-rhodamine}), and G\textsub{M1} ganglioside (GM1)
from Avanti Polar Lipids.
All chemicals were used as received.
Deionized water with \SI{18.2}{M\ohm \centi\meter} resistivity, obtained using a Millipore Filtration System (Milli-Q Gradient A10), was used in all experiments.

\subsection*{Membrane preparation}
The here employed Giant Unilamellar Vesicles with diameters ranging from 5 to \SI{20}{\um} were produced using the standard electroswelling technique \cite{Angelova1986a}. A lipid mixture was prepared in chloroform at \SI{2}{g\per\liter} consisting of \SI{97.5}{wt.\percent} DOPC, \SI{2.0}{wt.\percent} DOPE-PEG-biotin, and \SI{0.5}{wt.\percent} DOPE-rhodamine. \SI{10}{\uL} of this solution was dried on each of two \SI{6}{cm^2} ITO-coated glass slides (\SIrange{15}{25}{\ohm}, Sigma-Aldrich) and subsequently submersed in a solution containing \SI{100}{mM} glucose and \SI{0.3}{mM} sodium azide in 49:51 D\textsub{2}O:H\textsub{2}O. To induce vesicle formation, the electrodes were subjected to \SI{1.1}{V} (rms) at \SI{10}{Hz} for \SI{2}{hour}. The thus obtained vesicles were stored in a BSA-coated vial at room temperature. To separate the GUVs from smaller lipid vesicles, we filtered the GUV solution using a Whatmann \SI{5.0}{\um} pore size cellulosenitrate filter directly before use.
For the membrane tubulation experiments, the outer membrane leaflet was stained using the ganglioside GM1 and the dye cholera toxin 488, as follows: the above lipid mixture in chloroform was transferred to methanol, adding \SI{0.2}{wt.\percent} GM1.
The electroswelling then proceeded as descibed above.
Before sample preparation, \SI{50}{mg\per\liter} cholera toxin 488 was added to the PBS to stain the outer leaflet of the vesicles only.

\subsection*{Microparticle preparation}
Carboxylated polystyrene microparticles (diameter \SI{0.98+-0.03}{\um}) were synthesized in a surfactant-free radical polymerization from \SI{25.0}{g} styrene, \SI{500}{mg} itaconic acid, \SI{250}{mg} ACVA, and \SI{125}{mL} water \cite{Appel2013}. \SI{10}{mg} of the fluorescent dye BODIPY was included during synthesis. The size distribution of the particles was measured from electron micrographs taken by an FEI NanoSEM 200.
To ensure a specific and strong adhesion between the microparticles and the lipid membrane we coated the microparticles with avidin and mPEG, using an EDC/Sulfo-NHS coating procedure \cite{Meng2004}. Per \SI{15}{mg} particles, \SIrange{15}{50}{\ug} NeutrAvidin and \SI{4.0}{mg} mPEG were used.
Sodium azide was added to a concentration of \SI{3}{mM} to prevent bacterial growth.
The density of biotin binding sites (\SIrange{5e1}{1.5e3}{\per\square\micro\meter}) was quantified using a fluorescence assay with DNA oligonucleotides having a biotin and a fluorescent marker. The particle coating as well as the quantification of biotin binding sites are described extensively in ref. \cite{VanderWel2016b}.

\subsection*{Sample preparation}
To attach the microparticles to the model membranes, we gently mixed \SI{20}{\micro \liter} of the GUVs with \SI{0.04}{\micro \gram} microparticles dispersed in \SI{6}{\uL} phosphate buffered saline solution (PBS, pH=7.4). The PBS was density-matched with the microparticles (using D\textsub{2}O) and had the same osmolarity as the glucose inside the GUV. After an incubation time of \SI{10}{\minute}, the GUV-particle mixture was distributed into a microscope sample holder containing \SI{50}{\micro \liter} of PBS. Due to the small density difference between PBS and glucose solutions, the particle-covered GUVs sediment to the bottom and could be visualized with an inverted microscope.

In order to control the tension of the vesicles, we change the osmotic pressure outside of the vesicles as follows: If the sample holder is sealed directly after preparation, there is no water evaporation and the osmotic pressure is preserved. In this case, no fluctuations of the membrane are observed and the vesicles are denoted ``tense'', which for the here employed DOPC membranes corresponds to a membrane tension that is above \SI{1}{\micro \newton \per \meter}. If the sample is imaged while leaving the sample holder open for a period of \SIrange{30}{60}{\minute}, water evaporates, the osmolarity of the outside solution increases, and the membrane tension of the vesicles decreases. During this process, we are able to image so-called ``floppy'' vesicles that exhibit membrane undulations with an amplitude greater than \SI{1}{\um}. The corresponding membrane tension that follows from the fluctuation analysis described in ref. \cite{Pecreaux2004} lays below \SI{0.2}{\micro \newton \per \meter}. Figure~S1 shows such a ``floppy'' vesicle together with its fluctuation spectrum.

As described in ref. \cite{VanderWel2016b}, the microparticles attach spontaneously to the model membranes due to the strong non-covalent binding between NeutrAvidin and the biotinylated lipids. If the particle-membrane adhesion is larger than the membrane bending energy, and the vesicle is floppy, particles will be wrapped by the membrane, which is observed by the colocalization of membrane and particle fluorescence. As the membrane-particle linkage is irreversible, tense membranes with wrapped particles can be obtained by increasing the tension of an initially floppy membrane, for instance by wrapping of additional particles.

\subsection*{Diffusion measurements}
Diffusion measurements of membrane-attached particles were performed by imaging the top part of a tense GUV with attached particles (see Figure~S2). The particles were tracked in the recorded video images using Trackpy \cite{Allan2015}. Simultaneously, the fluorescence signal of the top part of the vesicle was recorded. We extract the center of the vesicle from a two-dimensional Gaussian fit of this fluorescence. Using a separate measurement of the vesicle size, the full three-dimensional coordinates of the particles could be reconstructed.

Diffusion measurements of wrapped particles were performed on sequences of confocal slices of particle-covered GUVs (see Figure~S3). Wrapped particles were identified by colocalization of membrane and particle fluorescence. The particles were tracked in the same way as with the attached particles. The vesicle position was determined by fitting circles to the vesicle contour \cite{VanderWel2016}. The particle displacements were then measured in one dimension along the vesicle contour.

Drifting vesicles were not analyzed to rule out particle drift due to vesicle rotation.
Particles that were closer than \SI{2.5}{\um} to other particles were also omitted to rule out the long-ranged interparticle forces caused by membrane deformation \cite{Sarfati2016a,VanderWel2016b}.
The diffusion coefficient was measured for each particle separately with linear regression of the mean squared displacements.
As particles were confined to a spherical surface, the mean squared displacement only grows linearly in time if the squared curvature $R^2$ is much larger than the measured mean squared displacement \cite{Paquay2016a}.
In order to meet this condition, we limited the displacement measurements to short time intervals up to 4 frames, corresponding to maximum lag times of \SIrange{35}{70}{ms}.

The measurement precision of each single-particle diffusion coefficient depends on the measured trajectory length \cite{Qian1991, Ernst2013}.
We computed these uncertainties and omitted measurements with an imprecision larger than \SI{0.04}{\square\um\per\second} from the histograms of single-particle diffusion coefficients.
This resulted in 39 trajectories (8 different GUVs) of attached particles and 25 trajectories of wrapped particles (on 14 different GUVs).

\subsection*{Imaging}
Samples were imaged with a Nikon Ti-E A1R confocal microscope equipped with a $60\times$ water immersion objective (NA = 1.2). The BODIPY and cholera toxin 488 (ex. \SI{488}{nm}, em. \SI{525+-25}{nm}) and DOPE-rhodamine (ex. \SI{561}{nm}, em. \SI{595+-30}{nm}) emissions were recorded simultaneously with a dichroic mirror separating the emission signal onto two detectors. The coverslips were pre-treated with a layer of polyacrylamide to prevent particle and GUV adhesion, using a method described in ref. \cite{VanderWel2016b}. High-speed images used in the diffusion measurements were recorded at \SIrange{57}{110}{Hz} using a horizontal resonant scanning mirror, while high-resolution close-ups were recorded with a set of Galvano scanning mirrors.

\section*{Results and Discussion}
Polymer microparticles adhere to lipid membranes by various kinds of (bio)chemical interactions, such as Van der Waals forces, electrostatic forces, or complementary protein interactions \cite{Nel2009, Mahmoudi2014, Vonarbourg2006a}.
In our experiments, we realized particle adhesion in a controlled manner by coating polystyrene particles (\SI{1}{\um} diameter) with the protein avidin and including a biotinylated lipid in the membranes.
Once connected to the membrane, the particles do not detach due to the high energy gain associated with the non-covalent linkage between biotin and avidin (approx. $\SI{17}{k_BT}$ \cite{Moy1994}, with $\mathrm{k_BT}$ being the thermal energy), and the possibility of forming multiple avidin-biotin connections. An individual membrane-associated sphere usually occupies only one of two equilibrium states: either the membrane fully wraps around the microparticle or it does not deform at all \cite{Lipowsky1998a, Lipowsky1998, Raatz2014a, Agudo-Canalejo2015a}.
This can be tuned by changing the membrane tension and the membrane-particle adhesion energy \cite{VanderWel2016b}.

In this article, we will investigate the dynamics of particles in the attached and the wrapped state as well as their assembly behavior over time.
We will first probe the membrane association of isolated particles by measuring the influence of the membrane on the particle mobility.
Then, we will map out the assembly pathways of multiple particles that interact through membrane deformations and eventually lead to membrane-mediated aggregates.
Finally, we will investigate a spontaneous tubulation process that occurs at particle inclusions. 

\subsection*{Mobility of membrane-associated particles}
While a particle retains its lateral mobility when it adheres to a liquid lipid membrane, the Brownian motion of the particle changes from three-dimensional to two-dimensional.
Therefore its mean squared displacement changes from $\langle x^2 \rangle = 6Dt$ in the freely dispersed case to $4Dt$ in the bound case, where $D$ is the diffusion coefficient and $t$ denotes the lag time over which $\langle x^2 \rangle$ is measured. Additionally, upon particle adhesion we expect that its diffusion coefficient is lowered significantly due to an additional drag force caused by the relatively high viscosity of the membrane.

\begin{figure*}[h!]
	\centering{\includegraphics{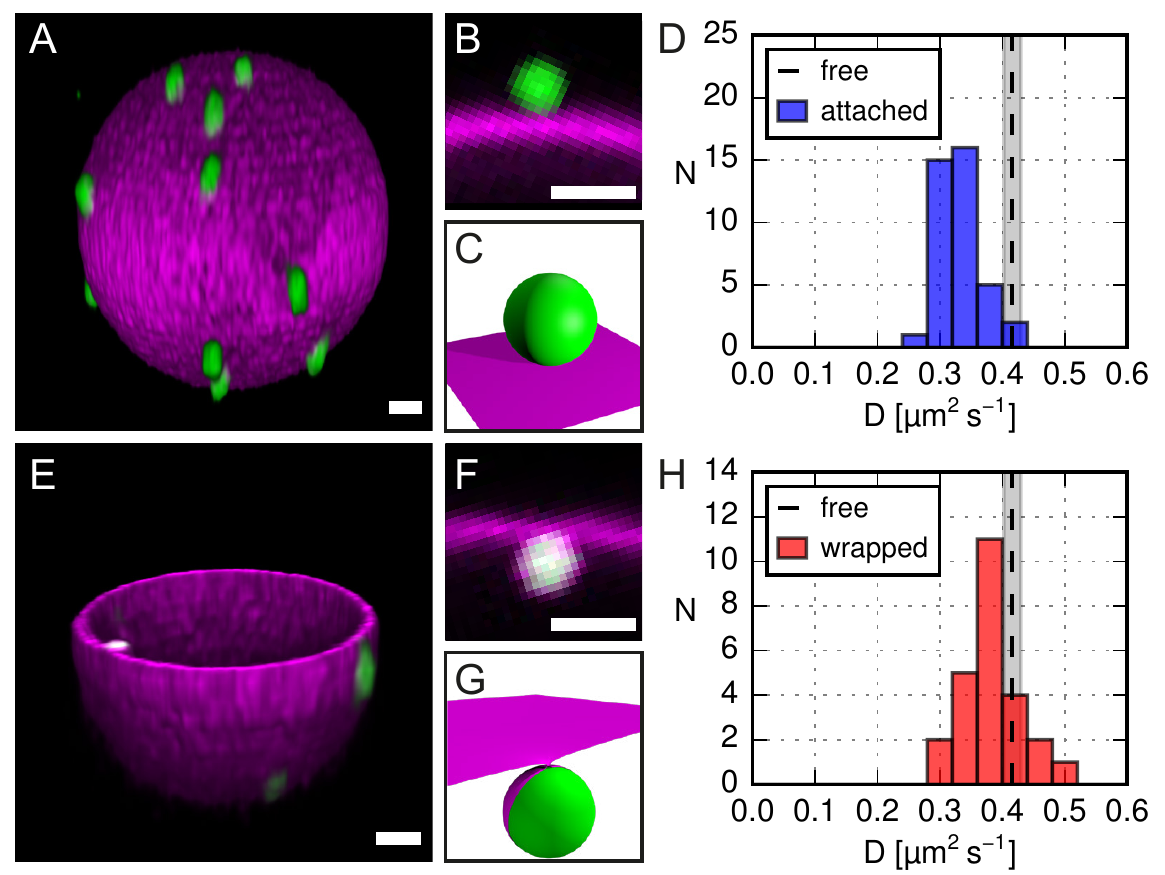}}
	\caption{Diffusion coefficients of membrane-adhered particles. In (\textit{A}), a three-dimensional reconstruction of a confocal image with particles attached to a vesicle is shown (scalebar \SI{2}{\um}), with the corresponding close-up in (\textit{B}) and illustration in (\textit{C}). A histogram of single-particle diffusion coefficients of these attached particles is shown in (\textit{D}). In (\textit{E}),
	one hemisphere of a membrane with one wrapped and two attached particles is displayed (scalebar \SI{2}{\um}), with (\textit{F}) the corresponding close-up, (\textit{G}) an illustration, and (\textit{H}) the histogram of single-particle diffusion coefficients. In the histograms, only coefficients with a standard error below \SI{0.04}{\square\um\per\second} were included. The binning width equals this value. For reference, the diffusion coefficient of freely dispersed particles is displayed by a dashed black line, with a shaded area that denotes the corresponding spread.}\label{fig:diffusion}
\end{figure*}

To quantify the the diffusion of attached particles, we used particles with biotin binding site densities ranging from \SI{5e1} to \SI{1.5e3}{\per\square\micro\meter} adhered to GUVs with diameters above \SI{15}{\um} and with a surface tension above \SI{1}{\micro \newton \per \meter}. This choice of membrane tension and biotin binding site density ensured that particles remained not wrapped (see Figure \ref{fig:diffusion}\textit{A}). As a reference, we first measured the diffusion constant of freely suspended particles and found it to be \SI{0.416+-0.014}{\square\um\per\second}, which is in agreement with the Stokes-Einstein relation.
The spread was determined from the known particle size distribution, which was measured from electron micrographs.
The distribution of the single-particle diffusion coefficients after membrane attachment is shown in Figure \ref{fig:diffusion}\textit{D}.
Compared to freely suspended particles, the diffusion coefficient decreased significantly to \SI{0.33+-0.04}{\square\um\per\second}.
The corresponding increase in drag coefficient $\zeta = k_BT / D$ due to the proximity of the membrane is 26\%.
Next to the measurement imprecision, this apparent spread of single-particle diffusion coefficients may also be caused by different avidin coating densities, yielding different particle-membrane adhesion shapes.

The diffusion measurements of wrapped particles were performed using a single particle batch with a higher biotin binding site density ranging between \SIrange{1.3e2}{1.5e3}{\per\square\micro\meter}.
To ensure a wrapped fraction of particles of 60\%, we increased the available membrane surface area by lowering the tension of the GUVs temporarily below \SI{10}{\nano \newton \per \meter} \cite{VanderWel2016b}.
The resulting histogram of single-particle diffusion coefficients is shown in Figure \ref{fig:diffusion}\textit{H}.
Again the membrane association lowered the diffusivity of the particles, albeit less than for simple membrane attachment: we found \SI{0.38+-0.05}{\square\um\per\second} as the diffusion coefficient of wrapped particles.
The corresponding increase in drag is only 9\%.
Apparently, the membrane has less effect on the particle's mobility if it wraps around the particle.
We hypothesize that this can be explained by the membrane geometry around a wrapped particle.
As the membrane requires room to bend away from the particle, there is a finite distance between the particle and bulk membrane.
The farther the particle from the membrane, the lower the influence of the membrane on the particle \cite{Dimova1999a} and therefore this effect would result in a diffusion coefficient which is larger than in the non-wrapped case described above.

To quantitatively explain the excess drag exerted by a membrane on a Brownian particle, several models have been proposed relating the membrane surface viscosity to the particle's drag coefficient.
As the employed spheres of \SI{1}{\um} diameter are much larger than the membrane thickness, we cannot use models based on the Saffman-Delbr\"uck description \cite{Saffman1975,Hughes1981,Petrov2008}, which only consider the drag experienced by a circular membrane patch, and not by the full sphere that is attached to it.
This is seen conveniently by evaluating the dimensionless Boussinesq number $B = \eta_S / (\eta R)$, with $\eta_S$ the membrane surface viscosity, $\eta$ the bulk viscosity, and $R$ the sphere radius.
For the Saffman-Delbr\"uck model to be applicable, bulk viscous effects on the sphere have to be negligible and $B \gg 1$.
We estimate $B = 10$ for our system and therefore employ the analysis by Dimowa et al, which is a numerical solution to the full hydrodynamic model for spheres embedded in a viscous membrane \cite{Danov1995}.
As they found the counter-intuitive result that the drag coefficient only slightly depends on the inclusion height of the particle relative to the membrane, we may invoke their numerical relation without precisely knowing the particle-membrane adhesion patch size.
From our measured excess drag coefficient we thereby obtain $\eta_S = \SI{1.2+-0.1e-9}{\pascal \second \meter}$.
This DOPC membrane viscosity lies in between previously reported values of \SI{5.9+-0.2e-10}{Pa s m} \cite{Herold2010} and \SI{15.9+-2.3e-9}{Pa s m} \cite{Hormel2014}.

To summarize, we observed that membrane-attached particles experience an excess drag of 26\% due to the membrane.
Using the model by Dimova and Danov, and neglecting differences due to inclusion height, we obtained a membrane viscosity of \SI{1.2+-0.1e-9}{Pa s m}.
Furthermore, we observed that membrane-wrapped particles only experience an excess drag of 9\%, which we explain by a finite distance between membrane and particle which decreases the influence of the membrane viscosity.
The here described diffusion measurements have direct consequences for the timescales of the diffusion-limited aggregation processes that are described in the next section.

\subsection*{Microparticle assembly pathways}
\begin{figure*}[h!]
	\centering{\includegraphics{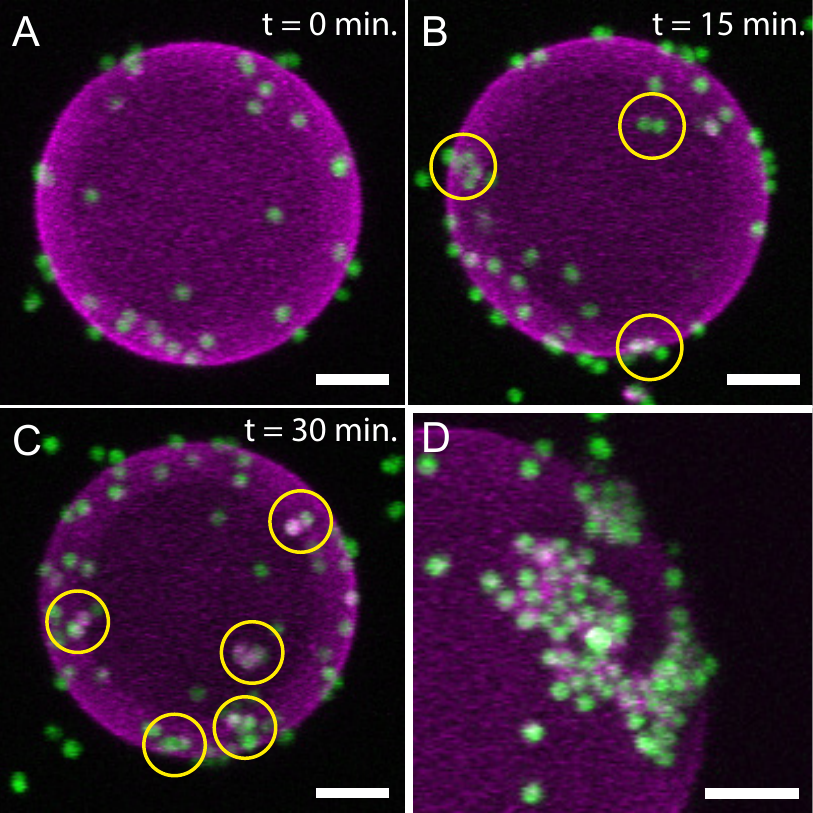}}
	\caption{Microparticles that aggregate on a lipid vesicle. A vesicle (in magenta) with associated particles (in green) is shown (\textit{A}) directly after sample preparation, (\textit{B}) after \SI{15}{\minute}, and (\textit{C}) after \SI{30}{\minute}. The images are maximum intensity projections. Formed aggregates, which were identified from the full three-dimensional images, are denoted by yellow circles. (\textit{D}) Through membrane-mediated aggregation, particle clusters may grow to include several tens of particles. (\textit{D}) was recorded in a separate experiment under the same conditions. Scale bars are \SI{5}{\um}.}
	\label{fig:clustering}
\end{figure*}

\begin{figure*}[t]  
	\centering{\includegraphics{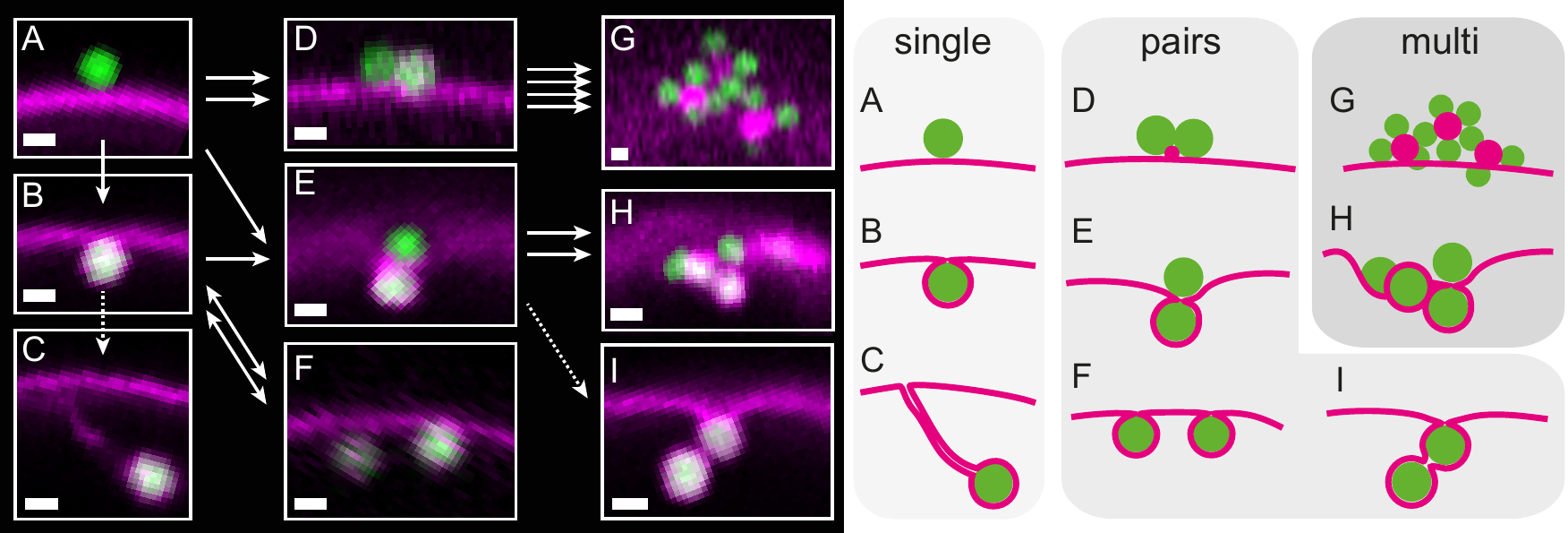}}
	\caption{Membrane-mediated irreversible assembly pathways of microparticles, displayed in confocal images (left, scale bars: \SI{1}{\um}) and sketches (right). Solid arrows denote irreversible particle assembly pathways, two-sided arrows denote a reversible pathway, and dotted arrows denote pathways that were observed less than 1\% of the cases. \textit{(A)} A membrane-attached particle is \textit{(B)} wrapped at large membrane-particle adhesion energy and low membrane tension. \textit{(C)} A wrapped particle can induce the formation of a membrane tube (see also Movie~S1).  Subsequently, \textit{(D)} two attached particles stick together via secondary lipid vesicles.
	\textit{(E)} An attached particle assembles with a wrapped particle forming a membrane-mediated dimer (see also Movie~S2), while
	\textit{(F)} two wrapped particles interact reversibly \cite{VanderWel2016b}.
	\textit{(G)} Multiple particles form an aggregate via secondary lipid vesicles.
	\textit{(H)} The dimers aggregate into tetramers and larger structures (see also Movie~S3). 
	\textit{(I)} The membrane-mediated dimer \textit{(E)} can be fully wrapped by the membrane.
	The wrapped half of the dimer can be identified by the white color which is caused by the overlap of the membrane and particle fluorescence.
}\label{fig:pathways}
\end{figure*}

When particles were incubated with floppy vesicles for more than \SI{30}{\minute}, we observed the formation of large particle aggregates on the membrane.
For example, in Figure \ref{fig:clustering}, snapshots of the same vesicle at different points in time are shown, which demonstrates the gradual appearance of particle clusters through membrane-mediated aggregation.
The same particles were observed to not aggregate in the absence of vesicles: the assembly pathways described here only occur after adhesion to the membrane.
We found that the two states of particle-membrane adhesion, attached to and wrapped by the membrane (Figures \ref{fig:pathways}\textit{A} and \textit{B}), induce distinct assembly pathways, leading to the observed aggregation of particles on GUVs.
In this section, we will categorize and discuss their interactions and assembly pathways, starting from all combinations of single-particle adhesion states: (1) wrapped-wrapped, (2) attached-attached, and (3) wrapped-attached.

The first assembly pathway is due to the long ranged attraction exhibited by two membrane-wrapped particles (Figure \ref{fig:pathways}\textit{F}). This interaction force has been described extensively in theory and simulation \cite{Yolcu2014,Reynwar2011,Bahrami2012,Saric2012}, but only recently in experiment \cite{VanderWel2016b}. In the experimental system employed here, the attractive potential was found to range over \SI{2.5}{\um} with a strength of three times the thermal energy. For sparse coverage with microparticles, both assembly into a dimer state and subsequent disassembly were observed indicating reversibility of this assembly pathway. For a more dense coverage with wrapped particles, simulations predict the formation of linear and crystalline structures \cite{Dommersnes2002,Saric2011,Saric2012a}.

Second, membrane-attached particles (Figure \ref{fig:pathways}\textit{A}) do not interact with each other. However, in the presence of small ($R < \SI{1}{\um}$) lipid vesicles, aggregation of these particles can be observed \cite{VanderWel2016b}.
This leads to particle aggregates mediated by small lipid structures, as shown in Figures \ref{fig:pathways}\textit{D} and \textit{G}.
The small vesicles can only be identified by their fluorescence in confocal images, while they are typically invisible in bright field microscopy.
Contrary to the attached particles, fully wrapped particles cannot aggregate via small lipid vesicles, for the reason that their surface is already occupied by the GUV membrane.

\begin{figure}[t]  
	\centering{\includegraphics{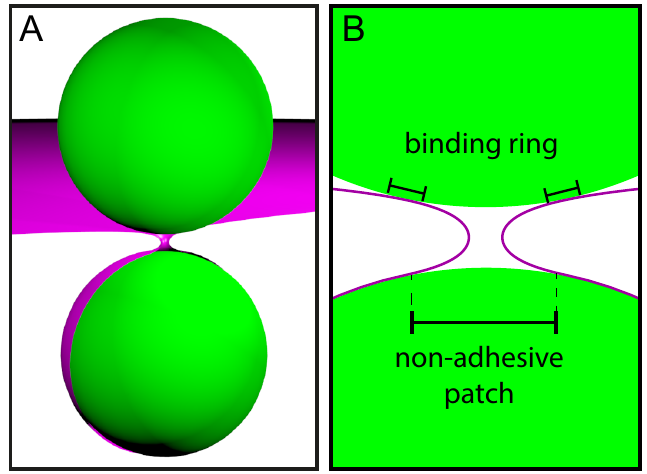}}
	\caption{Membrane-mediated dimer of microparticles. A three-dimensional illustration of the dimer is shown in \textit{(A)}. In \textit{(B)} a cross-section is depicted, showing the binding ring between the membrane and the top particle, which irreversibly connects the particles. The non-adhesive patch size was chosen arbitrarily for the purpose of this illustration; the membrane thickness is to scale with the particle diameter. See also Figure \ref{fig:pathways}\textit{E}.}\label{fig:dimer_render}
\end{figure}

A third and most complex pathway occurs for a combination of wrapped and attached particles and starts with a well-defined assembly step:
the local deformation of the membrane induced by the wrapped particles can serve as a binding site for attached particles (Figure \ref{fig:pathways}\textit{E}).
We observed this membrane-mediated dimerization only when an attached particle came into contact with a wrapped particle (see Movie~S2), and the resulting dimer structure occurs frequently on membranes that initially had both attached and wrapped particles present.
The dimers remain stable over the course of the experiment, unless they assemble further into larger aggregates (for example, Figure \ref{fig:pathways}\textit{H} and Movie~S3).
We hypothesize that the attached particles are captured irreversibly on top of the wrapped particle due to the increased contact area, as illustrated in Figure \ref{fig:dimer_render}.
Although the membrane-particle linkages are mobile in the membrane, the attached particle cannot escape once sufficient linkages have been formed in this ring-shaped region.
This is because the formed ring of bound linkers would need to cross the central non-binding patch, which requires the breakage of linkages.
In this way, the attached particle is topologically protected from detaching from the wrapped particle and they diffuse together as a membrane-mediated dimer.

Rarely, we observed a two-step hierarchical wrapping: after formation of the membrane-mediated dimer, the previously only attached particle becomes wrapped as well, resulting in a tube-like structure containing two wrapped spheres (see Figure \ref{fig:pathways}\textit{I}).
This indicates that the biotin binding site density on this particle is large enough to ultimately become wrapped but, coincidentally, this only occurred after it had been captured by the earlier wrapped particle. 
Although similar configurations of multiple particles wrapped in membrane tubes have been observed in simulations \cite{Saric2012,Bahrami2012,Raatz2014a}, it is important to note here that membrane-particle binding in our experiments is an irreversible process due to the strong avidin-biotin bonds.
Therefore, equilibrium considerations do not necessarily apply to our adhesion-driven interaction pathways.
Also, this implies that the tubularly wrapped spheres cannot be formed by the combination of two wrapped particles, as this would require partial unwrapping and thus breakage of membrane-particle linkages.
Yet, the presence of membrane deformation induced by the initially wrapped particle may further wrapping of the second particle by lowering a kinetic barrier towards the energetically more favorable wrapped state.

The relative frequencies of the three described dimerization pathways are fully determined by the initial particle wrapping state: we exclusively observe small-vesicle mediated interaction for attached particles, long ranged attraction for wrapped particles, and irreversible dimerization for the combination attached-wrapped. Thus, the short-ranged membrane-particle adhesion mechanism drives the lipid vesicle mediated interactions that result in aggregation of attached particles (Figure \ref{fig:pathways}\textit{D}) as well as in the dimerization of attached and wrapped particles (Figure \ref{fig:pathways}\textit{E}).

These adhesion-driven interactions continue to play a role as long as there is area on the particle left that is not covered by membrane, such as is the case for the attached particles (Figure \ref{fig:pathways}\textit{A} and \textit{D}) or for the membrane-mediated dimer (Figure \ref{fig:pathways}\textit{E}).
Indeed, two dimers can assemble into larger structures such as tetramers (Figure \ref{fig:pathways}\textit{H}), and multiple attached particles aggregate via secondary lipid vesicles (Figure \ref{fig:pathways}\textit{G}).
Depending on the particle concentration and waiting time after mixing, particles were observed to assemble on floppy membranes into permanent aggregates of 5-50 particles (see Figure \ref{fig:clustering}\textit{D}). 
On tense membranes, however, particles do not become wrapped and therefore only the small lipid vesicle mediated aggregation can be observed (Figures \ref{fig:pathways}\textit{D} and \textit{G}).

In summary, by using confocal microscopy we were able to distinguish a variety of assembled microparticle structures mediated purely by a lipid membrane, either through membrane deformations or the short-ranged adhesive forces between membrane and particles.
We correlated the different observed structures with the initial state of wrapping of the individual particles and formulated an aggregation model containing three mechanisms for membrane-mediated interactions.
Apart from the secondary lipid structure interactions, which can be suppressed \cite{VanderWel2016b}, we can categorize collective particle interactions on membranes into three different types:
First, membrane-attached particles exhibit only diffusive motion and do not interact with each other.
Second, membrane-wrapped particles interact with each other via membrane deformation forces. Multiple wrapped particles are expected to form linear and crystalline structures \cite{Dommersnes2002,Saric2011, Saric2012a}.
However, we did not observe this behavior because of the third regime: a mixture of attached and wrapped particles starts with the formation of permanent dimers and eventually leads to permanent particle aggregates. 
This observation of microparticle aggregation on GUVs proves that for the formation of membrane-mediated aggregates, no cytoskeleton or other active components are necessary: the adhesion between particle and membrane is sufficient for aggregation.

When considering the impact of microparticles on living organisms, internalization into the cell has been the major focus in literature, although aggregation on cellular membranes has been reported \cite{Rejman2004, Lorenz2006, Gaumet2009}.
We here showed that this aggregation is driven by the membrane itself and therefore is a generic property of microparticles that adhere to cellular membranes.
This has the following important consequence on the particle retention timescales: whereas a single microparticle may be bound reversibly to the membrane, aggregates will form multiple bonds with the membrane which take exponentially longer times to break \cite{Gunnarsson2011}.
Therefore, the here described membrane-mediated aggregation of microparticles might be an important factor contributing to the high retention times of polymer microparticles in organisms.

\subsection*{Spontaneous membrane tubulation}
\begin{figure*}[t]  
	\centering{\includegraphics{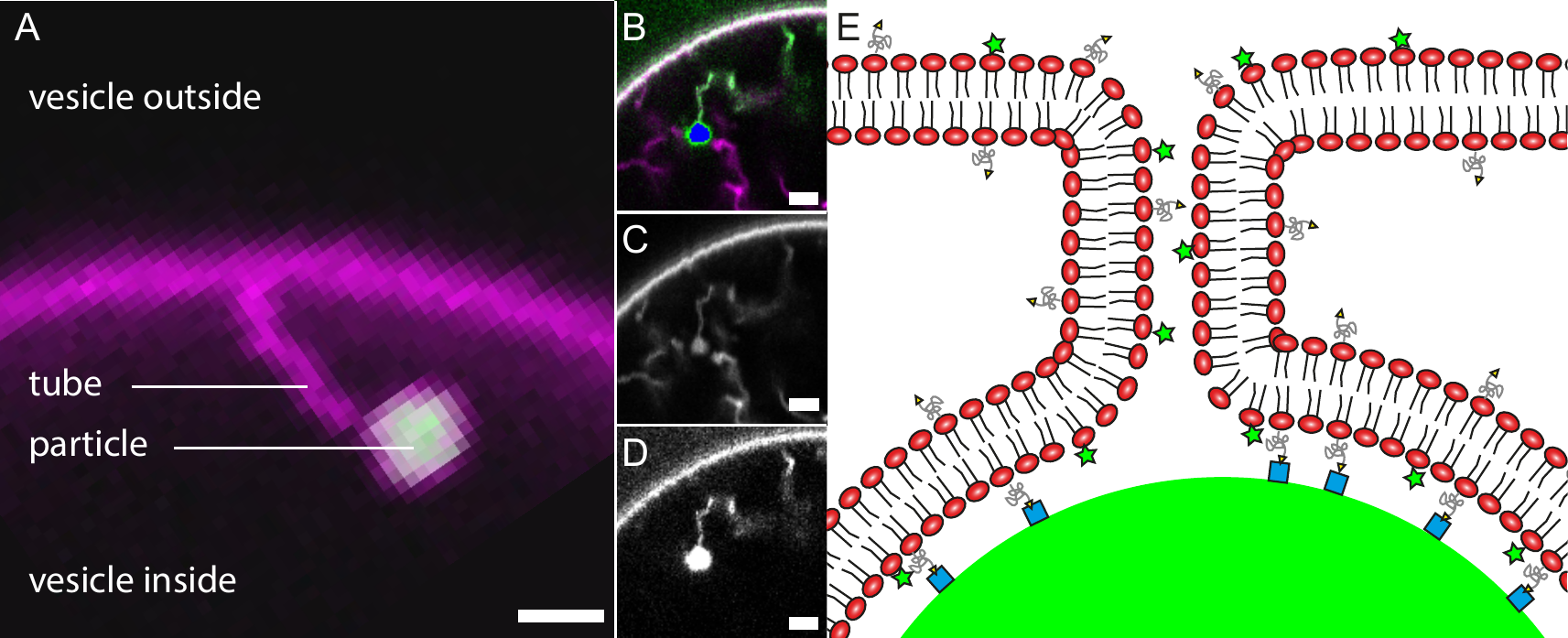}}
	\caption{Spontaneous membrane tubulation at wrapped microparticles.
	(\textit{A}) A confocal close-up of a particle (green) that is connected to the membrane (magenta) via a tube. See also corresponding Movie~S1.
	In order to stain only the outer membrane leaflet, we included GM1 lipids into the GUVs and added cholera toxin 488 after GUV formation.
	(\textit{B}) A channel overlay with the membrane in magenta, cholera toxin 488 in green, and the particle fluorescence in blue.
	Single channel images in the wavelength range \SIrange{500}{550}{nm} (\textit{C})  shows only the DOPE-rhodamine, and \SIrange{565}{625}{nm} (\textit{D}) shows the cholera toxin 488 together with the microparticle.
	As the particle-induced tube clearly exhibited green fluorescence, we conclude that the outer leaflet is present and therefore the structure must be a membrane tube.
	The absence of green fluorescence in the other membrane structures inside the GUV confirms that the cholera toxin 488 only stains the vesicle outside.	
	(\textit{E}) An illustration of the membrane tube. Here, green stars denote the cholera toxin 488 which is attached to GM1, yellow triangles the biotin which is connected via a PEG2000 spacer to DOPE lipids, and blue squares the NeutrAvidin which is connected to the particle. Scalebars denote \SI{2}{\um}.}\label{fig:tubulation}
\end{figure*}

After a particle had been wrapped by the lipid membrane, we occasionally observed ($N=20$) spontaneous formation of a membrane tube starting from the neck of the wrapped particle onwards (see Figure \ref{fig:tubulation}\textit{A} and Movie~S1).
This tubulation process occurred for less than 1\% of all wrapped particles.
The particles remained wrapped by and connected to the vesicle membrane, but could freely diffuse in the vesicle interior. The membrane tubes elongated rapidly and exhibited large thermal fluctuations. In addition, we never observed reversal of the tubulation, which underlines the surprising absence of retracting tensile forces.

In order to prove that the observed structure is in fact a membrane tube, and not a rodlike micelle, we enable distinction between the inner and outer leaflet of the membrane through inclusion of a fluorescent dye.
By adding choleratoxin 488 after formation of GUVs that contained \SI{0.2}{wt.\percent} GM1, we achieved exclusive staining of the outer leaflet.
This is confirmed by the absence of green fluorescence in other membrane structures inside the GUV (see Figure \ref{fig:tubulation}\textit{B-D}).
If the particle would have induced hemifusion of the membrane thereby creating a tubular micelle that connected the wrapped particle to the membrane, the tube would consist of the inner leaflet only.
Conversely, the membrane would be made up of both the inner and outer leaflet if the particle had truly produced a tubular structure.
The fluorscence of choleatoxin 488 in the tube observed in the experiment clearly indicated that the outer leaflet is present.
Thus, we can conclude that the observed structure in between the particle and the GUV in fact consists of a strongly curved tubular membrane.

Why do membrane tubes spontaneously nucleate at wrapped particles?
While this process is ubiquitous in living cells \cite{Terasaki1986,Hopkins1990,Heinrich2014}, it is typically not observed in (symmetric) model bilayers.
In the absence of preferred curvature, it has been established both theoretically and experimentally that a supporting force $F = 2 \pi \sqrt{2\sigma\kappa}$ is necessary to stabilize a tube against retraction, and that its radius is given by $R = \sqrt{\kappa / (2\sigma)}$  \cite{Derenyi2002, Koster2005, Shaklee2008}.
Here, $\sigma$ denotes the membrane tension and $\kappa$ its bending rigidity.
Taken together, this yields $F = 2 \pi \kappa / R$.
Since the membrane diameter of our observed tubes is below the diffraction limit, $R < \SI{100}{nm}$, a tube-supporting tensile force would need to be larger than \SI{5}{pN} (given $\kappa = \SI{21}{k_BT}$ \cite{Rawicz2000}).
Clearly, such a force is absent in our experiments, as the particle diffuses freely after tube-formation for the duration of the experiment. A force of \SI{1}{pN} on a freely suspended colloidal particle would already result in a particle drift velocity of \SI{100}{\um \per \second}.
Thus, we conclude that only a preferred curvature of the bilayer could explain the observed membrane tubes \cite{Li2011}.

The bulky DOPE-PEG-biotin lipids could indeed generate such a preferred curvature, if they would be distributed unevenly across the bilayer leaflets \cite{Lipowsky2013}.
Depletion of these lipids by adhesion to the particles seems however unlikely: the here employed DOPE-PEG-biotin molar fraction of 0.5\% is well above the number of biotin binding sites on the particles of linkers on the particles.
Therefore, an initial bilayer asymmetry due to an uneven distribution of lipids or due to solvent asymmetry \cite{Dobereiner1999} are likely causes of the observed preferred bilayer curvature.

Given the presence of a preferred curvature, the neck that is induced by the wrapped particles logically serves as a starting point for the formation of a tube, which would otherwise have to cross a significant energy barrier \cite{Koster2005}.
In this way, membrane-wrapped microparticles act as nucleation sites for membrane tubes.

\section*{Conclusion}
We have shown that microparticles aggregate on lipid membranes.
After adhesion to the membrane, particles remain laterally mobile on the membrane. We have quantified the particles' mobility and found that their diffusion coefficients of \SI{0.416+-0.014}{\square\um\per\second} in dispersion are reduced to \SI{0.33+-0.04}{\square\um\per\second} after membrane attachment.
Using the numerical model by Dimova and Danov \cite{Danov1995,Dimova1999a,Danov2000}, we computed the membrane surface viscosity and found \SI{1.2+-0.7e-9}{Pa s m}.
This value falls in between two previously determined values for a pure DOPC membrane \cite{Herold2010, Hormel2014}.
For wrapped particles, we observed diffusion coefficients of \SI{0.38+-0.05}{\square\um\per\second}.
We hypothesize that the higher diffusivity is due to an increased distance between membrane and particle.

We have established a complete picture of the onset of aggregation of microparticles associated with lipid membranes.
Particles can interact with each other in the following three ways.
Firstly, membrane-wrapped particles interact with each other through a membrane deformation mediated force that ranges over several particle diameters, as has been shown before \cite{Koltover1999,Sarfati2016a,VanderWel2016b}.
Secondly, this interaction is not present for attached particles that do not deform the membrane.
However, these can irreversibly aggregate via smaller secondary vesicles \cite{VanderWel2016b}.
Thirdly, we have shown here that attached particles can become trapped in the deformation sites created by a wrapped particle, forming membrane-mediated dimers.
The last two adhesion-mediated mechanisms ultimately result in complex random aggregates of multiple particles.
In the case in which the membrane has a preferred curvature, wrapped microparticles can also serve as a nucleation site for membrane tubes.

The observed aggregation of colloidal particles mediated by lipid membranes shows that even in the absence of proteins or active components, particles aggregate in order to optimize their contact area with the lipid membrane, or to minimize the membrane deformation.
This has important consequences: whereas in general single microparticles may easily adsorb and desorb on the membrane, aggregates of particles desorb exponentially slower, which might explain the reported high retention times in organisms.
This systematic description of the assembly pathways of microparticles on model lipid membranes will enable a better understanding of microparticle aggregation in biological systems, and especially the accumulation of microplastics.

\section*{Author Contributions}
D.J.K. and D.H. designed the project; C.v.d.W. carried out experiments and data analysis supervised by D.J.K. and D.H.; all authors contributed to writing and revising the manuscript.

\section*{Acknowledgement}
The authors thank Timon Idema and Agur Sevink for useful discussions and Marcel Winter for support during particle synthesis. This work was supported by the Netherlands Organisation for Scientific Research (NWO/OCW), as part of the Frontiers of Nanoscience program and VENI grant 680-47-431.


\begin{thebibliography}{10}
	
	\bibitem{Tang2012}
	F.~Tang, L.~Li, and D.~Chen.
	\newblock {Mesoporous silica nanoparticles: Synthesis, biocompatibility and
		drug delivery}.
	\newblock {\em Adv. Mater.}, 24:1504--1534, 2012.
	
	\bibitem{Tseng2002}
	Y.~Tseng, T.~P. Kole, and D.~Wirtz.
	\newblock {Micromechanical mapping of live cells by multiple-particle-tracking
		microrheology}.
	\newblock {\em Biophys. J.}, 83(6):3162--3176, 2002.
	
	\bibitem{Cole2011a}
	M.~Cole, P.~Lindeque, C.~Halsband, and T.~S. Galloway.
	\newblock {Microplastics as contaminants in the marine environment: A review}.
	\newblock {\em Mar. Pollut. Bull.}, 62:2588--2597, 2011.
	
	\bibitem{Nel2009}
	A.~E. Nel, L.~M{\"{a}}dler, D.~Velegol, T.~Xia, E.~M.~V. Hoek, P.~Somasundaran,
	F.~Klaessig, V.~Castranova, and M.~Thompson.
	\newblock {Understanding biophysicochemical interactions at the nano-bio
		interface}.
	\newblock {\em Nat. Mater.}, 8:543--557, 2009.
	
	\bibitem{Mahmoudi2014}
	M.~Mahmoudi, J.~Meng, X.~Xue, X.~J. Liang, M.~Rahman, C.~Pfeiffer, R.~Hartmann,
	P.~R. Gil, B.~Pelaz, W.~J. Parak, P.~del Pino, S.~Carregal-Romero, A.~G.
	Kanaras, and S.~{Tamil Selvan}.
	\newblock {Interaction of stable colloidal nanoparticles with cellular
		membranes}.
	\newblock {\em Biotechnol. Adv.}, 32(4):679--692, 2014.
	
	\bibitem{Vonarbourg2006a}
	A.~Vonarbourg, C.~Passirani, P.~Saulnier, and J.~P. Benoit.
	\newblock {Parameters influencing the stealthiness of colloidal drug delivery
		systems}.
	\newblock {\em Biomaterials}, 27(24):4356--4373, 2006.
	
	\bibitem{Rejman2004}
	J.~Rejman, V.~Oberle, I.~S. Zuhorn, and D.~Hoekstra.
	\newblock {Size-dependent internalization of particles via the pathways of
		clathrin- and caveolae-mediated endocytosis}.
	\newblock {\em Biochem. J.}, 377:159--169, 2004.
	
	\bibitem{Lorenz2006}
	M.~R. Lorenz, V.~Holzapfel, A.~Musyanovych, K.~Nothelfer, P.~Walther, H.~Frank,
	K.~Landfester, H.~Schrezenmeier, and V.~Mail{\"{a}}nder.
	\newblock {Uptake of functionalized, fluorescent-labeled polymeric particles in
		different cell lines and stem cells}.
	\newblock {\em Biomaterials}, 27(14):2820--2828, 2006.
	
	\bibitem{Gaumet2009}
	M.~Gaumet, R.~Gurny, and F.~Delie.
	\newblock {Localization and quantification of biodegradable particles in an
		intestinal cell model: The influence of particle size}.
	\newblock {\em Eur. J. Pharm. Sci.}, 36:465--473, 2009.
	
	\bibitem{Ramos1999}
	L.~Ramos, T.C. Lubensky, N.~Dan, P.~Nelson, and D.~A. Weitz.
	\newblock {Surfactant-mediated two-dimensional crystallization of colloidal
		crystals}.
	\newblock {\em Science}, 286:2325--2328, 1999.
	
	\bibitem{Koltover1999}
	I.~Koltover, J.~O. R{\"{a}}dler, and C.~R. Safinya.
	\newblock {Membrane mediated attraction and ordered aggregation of colloidal
		particles bound to giant phospholipid vesicles}.
	\newblock {\em Phys. Rev. Lett.}, 82(9):1991--1994, 1999.
	
	\bibitem{Sarfati2016a}
	R.~Sarfati and E.~R. Dufresne.
	\newblock {Long-range attraction of particles adhered to lipid vesicles}.
	\newblock {\em Phys. Rev. E}, 94:012604, 2016.
	
	\bibitem{VanderWel2016b}
	C.~van~der Wel, A.~Vahid, A.~{\v{S}}ari{\'{c}}, T.~Idema, D.~Heinrich, and
	D.~J. Kraft.
	\newblock {Lipid membrane-mediated attractions between curvature inducing
		objects}.
	\newblock {\em Sci. Rep.}, 6:32825, 2016.
	
	\bibitem{Angelova1986a}
	M.~I. Angelova and D.~S. Dimitrov.
	\newblock {Liposome Electroformation}.
	\newblock {\em Faraday Discuss. Chem. Soc.}, 81:303--311, 1986.
	
	\bibitem{Appel2013}
	J.~Appel, S.~Akerboom, R.~G. Fokkink, and J.~Sprakel.
	\newblock {Facile one-step synthesis of monodisperse micron-sized latex
		particles with highly carboxylated surfaces}.
	\newblock {\em Macromol. Rapid Comm.}, 34:1284--1288, 2013.
	
	\bibitem{Meng2004}
	F.~Meng, G.~H.~M. Engbers, and J.~Feijen.
	\newblock {Polyethylene glycol-grafted polystyrene particles}.
	\newblock {\em J. Biomed. Mater. Res. A}, 70A(1):49--58, 2004.
	
	\bibitem{Pecreaux2004}
	J.~P{\'{e}}cr{\'{e}}aux, H.-G. D{\"{o}}bereiner, J.~Prost, J.-F. Joanny, and
	P.~Bassereau.
	\newblock {Refined contour analysis of giant unilamellar vesicles}.
	\newblock {\em Eur. Phys. J. E}, 13:277--290, 2004.
	
	\bibitem{Allan2015}
	D.~B. Allan, T.~A. Caswell, N.~C. Keim, and C.~van~der Wel.
	\newblock {Trackpy v0.3.0}.
	\newblock {\em Zenodo}, 34028, nov 2015.
	
	\bibitem{VanderWel2016}
	C.~van~der Wel.
	\newblock {Circletracking v1.0}.
	\newblock {\em Zenodo}, 47216, 2016.
	
	\bibitem{Paquay2016a}
	S.~Paquay and R.~Kusters.
	\newblock {A method for molecular dynamics on curved surfaces}.
	\newblock {\em Biophys. J.}, 110:1226--1233, 2016.
	
	\bibitem{Qian1991}
	H.~Qian, M.~P. Sheetz, and E.~L. Elson.
	\newblock {Single particle tracking Analysis of diffusion and flow in
		two-dimensional systems}.
	\newblock {\em Biophys. J.}, 60:910--921, 1991.
	
	\bibitem{Ernst2013}
	D.~Ernst and J.~K{\"{o}}hler.
	\newblock {Measuring a diffusion coefficient by single-particle tracking:
		statistical analysis of experimental mean squared displacement curves}.
	\newblock {\em Phys. Chem. Chem. Phys.}, 15:845--849, 2013.
	
	\bibitem{Moy1994}
	V.~T. Moy, E.-L. Florin, and H.~E. Gaub.
	\newblock {Intermolecular forces and energies between ligands and receptors}.
	\newblock {\em Science}, 266:257--259, 1994.
	
	\bibitem{Lipowsky1998a}
	R.~Lipowsky, H.-G. D{\"{o}}bereiner, C.~Hiergeist, and V.~Indrani.
	\newblock {Membrane curvature induced by polymers and colloids}.
	\newblock {\em Physica A}, 249:536--543, 1998.
	
	\bibitem{Lipowsky1998}
	R.~Lipowsky and H.-G. D{\"{o}}bereiner.
	\newblock {Vesicles in contact with nanoparticles and colloids}.
	\newblock {\em Europhys. Lett.}, 43(2):219--225, 1998.
	
	\bibitem{Raatz2014a}
	M.~Raatz, R.~Lipowsky, and T.~R. Weikl.
	\newblock {Cooperative wrapping of nanoparticles by membrane tubes}.
	\newblock {\em Soft Matter}, 10:3570--3577, 2014.
	
	\bibitem{Agudo-Canalejo2015a}
	J.~Agudo-Canalejo and R.~Lipowsky.
	\newblock {Critical particle sizes for the engulfment of nanoparticles by
		membranes and vesicles with bilayer asymmetry}.
	\newblock {\em ACS Nano}, 9(4):3704--3720, 2015.
	
	\bibitem{Dimova1999a}
	R.~Dimova, C.~Dietrich, A.~Hadjiisky, K.~Danov, and B.~Pouligny.
	\newblock {Falling ball viscosimetry of giant vesicle membranes: Finite-size
		effects}.
	\newblock {\em Eur. Phys. J. B}, 12:589--598, 1999.
	
	\bibitem{Saffman1975}
	P.~G. Saffman and M.~Delbr{\"{u}}ck.
	\newblock {Brownian motion in biological membranes}.
	\newblock {\em P. Natl. Acad. Sci. USA}, 72(8):3111--3113, 1975.
	
	\bibitem{Hughes1981}
	B.~D. Hughes, B.~A. Pailthorpe, and L.~R. White.
	\newblock {The translational and rotational drag on a cylinder moving in a
		membrane}.
	\newblock {\em J. Fluid Mech.}, 110:349, 1981.
	
	\bibitem{Petrov2008}
	E.~P. Petrov and P.~Schwille.
	\newblock {Translational diffusion in lipid membranes beyond the
		Saffman-Delbr{\"{u}}ck approximation}.
	\newblock {\em Biophys. J.}, 94:L41, 2008.
	
	\bibitem{Danov1995}
	K.~Danov, R.~Aust, F.~Durst, and U.~Lange.
	\newblock {Influence of the surface viscosity on the hydrodynamic resistance
		and suface diffusivity of a large Brownian particle}.
	\newblock {\em J. Colloid Interf. Sci.}, 175:36--45, 1995.
	
	\bibitem{Herold2010}
	C.~Herold, P.~Schwille, and E.~P. Petrov.
	\newblock {DNA condensation at freestanding cationic lipid bilayers}.
	\newblock {\em Phys. Rev. Lett.}, 104:148102, 2010.
	
	\bibitem{Hormel2014}
	T.~T. Hormel, S.~Q. Kurihara, M.~K. Brennan, M.~C. Wozniak, and
	R.~Parthasarathy.
	\newblock {Measuring lipid membrane viscosity using rotational and
		translational probe diffusion}.
	\newblock {\em Phys. Rev. Lett.}, 112:188101, 2014.
	
	\bibitem{Yolcu2014}
	C.~Yolcu, R.~C. Haussman, and M.~Deserno.
	\newblock {The effective field theory approach towards membrane-mediated
		interactions between particles.}
	\newblock {\em Adv. Colloid Interfac.}, 208:89--109, jun 2014.
	
	\bibitem{Reynwar2011}
	B.~J. Reynwar and M.~Deserno.
	\newblock {Membrane-mediated interactions between circular particles in the
		strongly curved regime}.
	\newblock {\em Soft Matter}, 7:8567, 2011.
	
	\bibitem{Bahrami2012}
	A.~H. Bahrami, R.~Lipowsky, and T.~R. Weikl.
	\newblock {Tubulation and aggregation of spherical nanoparticles adsorbed on
		vesicles}.
	\newblock {\em Phys. Rev. Lett.}, 109(18):188102, oct 2012.
	
	\bibitem{Saric2012}
	A.~{\v{S}}ari{\'{c}} and A.~Cacciuto.
	\newblock {Mechanism of membrane tube formation induced by adhesive
		nanocomponents}.
	\newblock {\em Phys. Rev. Lett.}, 109(18):188101, oct 2012.
	
	\bibitem{Dommersnes2002}
	P.~G. Dommersnes and J.-B. Fournier.
	\newblock {The many-body problem for anisotropic membrane inclusions and the
		self-assembly of "saddle" defects into an "egg carton"}.
	\newblock {\em Biophys. J.}, 83(6):2898--2905, 2002.
	
	\bibitem{Saric2011}
	A.~{\v{S}}ari{\'{c}} and A.~Cacciuto.
	\newblock {Soft elastic surfaces as a platform for particle self-assembly}.
	\newblock {\em Soft Matter}, 7:8324, 2011.
	
	\bibitem{Saric2012a}
	A.~{\v{S}}ari{\'{c}} and A.~Cacciuto.
	\newblock {Fluid membranes can drive linear aggregation of adsorbed spherical
		nanoparticles}.
	\newblock {\em Phys. Rev. Lett.}, 108:118101, 2012.
	
	\bibitem{Gunnarsson2011}
	A.~Gunnarsson, L.~Dexlin, P.~Wallin, S.~Svedhem, P.~J{\"{o}}nsson, C.~Wingren,
	and F.~H{\"{o}}{\"{o}}k.
	\newblock {Kinetics of ligand binding to membrane receptors from equilibrium
		fluctuation analysis of single binding events}.
	\newblock {\em J. Am. Chem. Soc.}, 133:14852--14855, 2011.
	
	\bibitem{Terasaki1986}
	M.~Terasaki, L.~B. Chen, and K.~Fujiwara.
	\newblock {Microtubules and the endoplasmic reticulum are highly interdependent
		structures}.
	\newblock {\em J. Cell Biol.}, 103:1557--1568, 1986.
	
	\bibitem{Hopkins1990}
	C.~R. Hopkins, A.~Gibson, M.~Shipman, and K.~Miller.
	\newblock {Movement of internalized ligand-receptor complexes along a
		continuous endosomal reticulum}.
	\newblock {\em Nature}, 346:335--339, 1990.
	
	\bibitem{Heinrich2014}
	D.~Heinrich, M.~Ecke, M.~Jasnin, U.~Engel, and G.~Gerisch.
	\newblock {Reversible membrane pearling in live cells upon destruction of the
		actin cortex}.
	\newblock {\em Biophys. J.}, 106(5):1079--1091, mar 2014.
	
	\bibitem{Derenyi2002}
	I.~Der{\'{e}}nyi, F.~J{\"{u}}licher, and J.~Prost.
	\newblock {Formation and interaction of membrane tubes}.
	\newblock {\em Phys. Rev. Lett.}, 88(23):238101, may 2002.
	
	\bibitem{Koster2005}
	G.~Koster, A.~Cacciuto, I.~Der{\'{e}}nyi, D.~Frenkel, and M.~Dogterom.
	\newblock {Force barriers for membrane tube formation}.
	\newblock {\em Phys. Rev. Lett.}, 94:068101, 2005.
	
	\bibitem{Shaklee2008}
	P.~M. Shaklee, T.~Idema, G.~Koster, C.~Storm, T.~Schmidt, and M.~Dogterom.
	\newblock {Bidirectional membrane tube dynamics driven by nonprocessive
		motors}.
	\newblock {\em P. Natl. Acad. Sci. USA}, 105(23):7993--7997, 2008.
	
	\bibitem{Rawicz2000}
	W.~Rawicz, K.~C. Olbrich, T.~McIntosh, D.~Needham, and E.~Evans.
	\newblock {Effect of chain length and unsaturation on elasticity of lipid
		bilayers}.
	\newblock {\em Biophys. J.}, 79:328--339, 2000.
	
	\bibitem{Li2011}
	Y.~Li, R.~Lipowsky, and R.~Dimova.
	\newblock {Membrane nanotubes induced by aqueous phase separation and
		stabilized by spontaneous curvature}.
	\newblock {\em P. Natl. Acad. Sci. USA}, 108(12):4731--4736, 2011.
	
	\bibitem{Lipowsky2013}
	R.~Lipowsky.
	\newblock {Spontaneous tubulation of membranes and vesicles reveals membrane
		tension generated by spontaneous curvature}.
	\newblock {\em Faraday Discuss.}, 161:305--331, 2013.
	
	\bibitem{Dobereiner1999}
	H.-G. D{\"{o}}bereiner, O.~Selchow, and R.~Lipowsky.
	\newblock {Spontaneous curvature of fluid vesicles induced by trans-bilayer
		sugar asymmetry}.
	\newblock {\em Eur. Biophys. J.}, 28:174--178, 1999.
	
	\bibitem{Danov2000}
	K.~D. Danov, R.~Dimova, and B.~Pouligny.
	\newblock {Viscous drag of a solid sphere straddling a spherical or flat
		surface}.
	\newblock {\em Phys. Fluids}, 12(11):2711--2722, 2000.
	
\end{thebibliography}
\end{document}


\section*{Supporting material for ``Microparticle assembly pathways on lipid membranes''}

\large Authors: C. van der Wel, D. Heinrich, and D.~J. Kraft

\section*{Supporting Videos}
\setcounter{figure}{0}
\makeatletter
\renewcommand{\fnum@figure}{Supporting Video S\thefigure}
\makeatother
Videos are separately available online. Here, still images of the videos are shown together with their captions.

\begin{figure}[H]
	\centering{\includegraphics[width=0.4\textwidth]{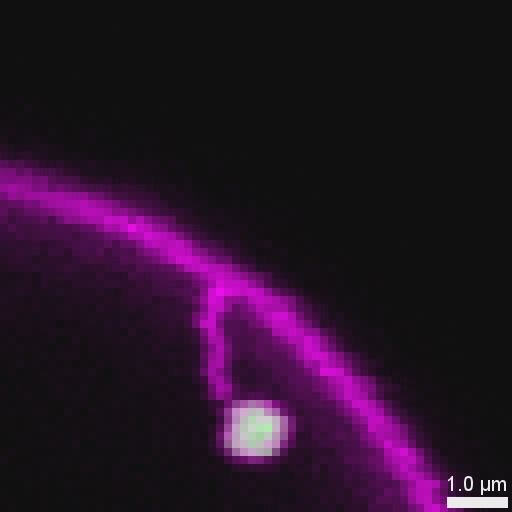}}
	\caption{Spontaneous tubulation induced by a \SI{1}{\um} colloidal particle wrapped by a lipid membrane. The movie shows an image sequence of confocal slices of a spherical vesicle (in magenta) with colloidal particles (in green), displayed in real time.}
\end{figure}

\begin{figure}[H]
	\centering{\includegraphics[width=0.4\textwidth]{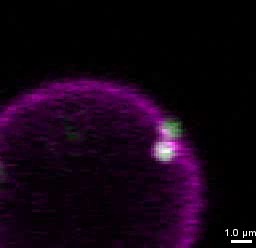}}
	\caption{Irreversible dimerization of a wrapped and attached particle. The movie shows an image sequence of confocal slices of a spherical vesicle (in magenta) with colloidal particles (in green), displayed in real time.
The wrapped particle has a white color due to the channel overlap. The attached particle is green. A (non-interacting) third attached particle is also present in the movie.}
\end{figure}

\begin{figure}[H]
	\centering{\includegraphics[width=0.4\textwidth]{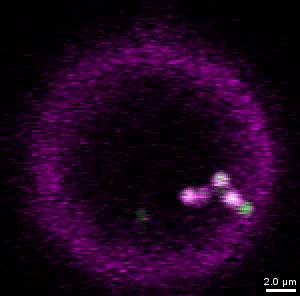}}
	\caption{Clustering of dimers into an aggregate of six particles. The movie shows an image sequence of confocal slices of a spherical vesicle (in magenta) with colloidal particles (in green), displayed in real time.
	One half of each dimer is depicted in white due to the overlap of the green and magenta fluorescence channels.}

\end{figure}

\section*{Supporting Figures}
\setcounter{figure}{0}
\makeatletter
\renewcommand{\fnum@figure}{Supporting Figure S\thefigure}
\makeatother

\begin{figure}[H]
	\centering{\includegraphics{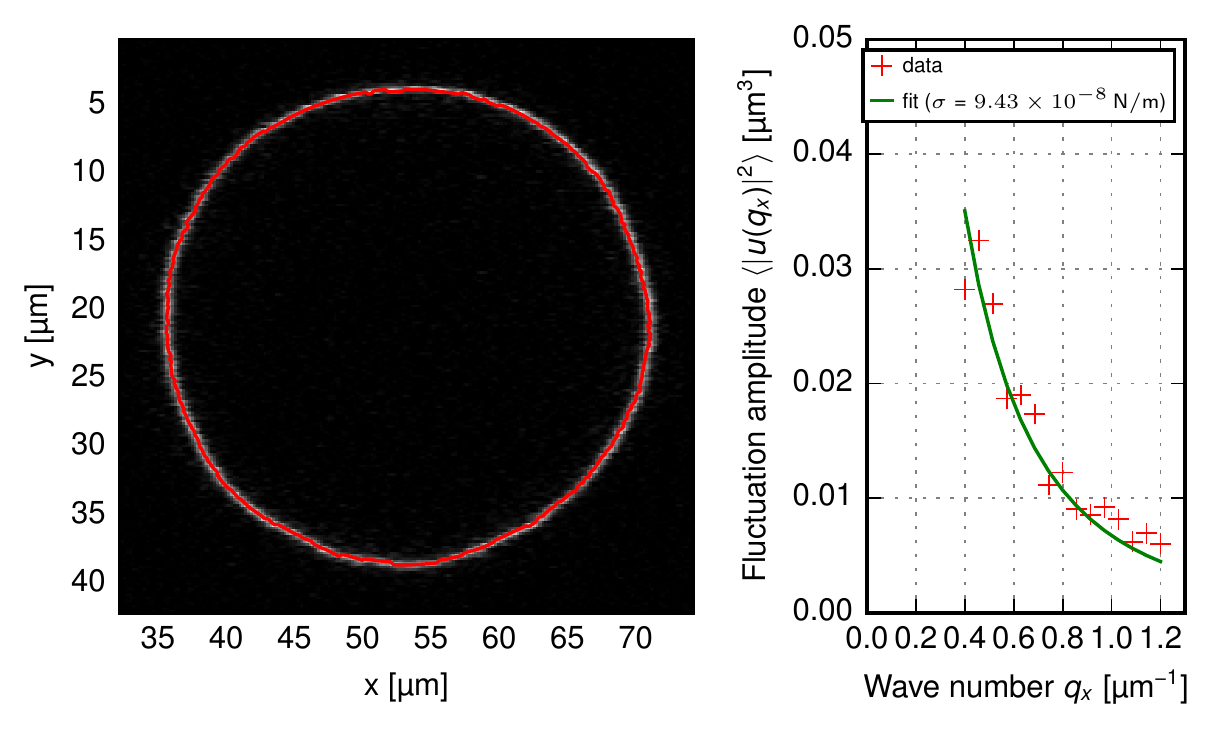}}
	\caption{Fluctuation analysis of a GUV. On the left, a cross-section of the vesicle is shown with its tracked profile as a red line. On the right, the amplitude of the discrete Fourier transform is shown (red crosses), together with a least-squares one-parameter fit to the theoretical power spectrum (green solid line). The power spectrum was averaged over 75 frames. The first five modes were discarded (corresponding to $q_x < \SI{0.4}{\um^{-1}}$) as the fit model is erroneous for these modes due to the closed topology of the vesicle. Also, wave numbers larger than \SI{1.2}{\um^{-1}} were omitted so that the correlation time of the fluctuations was at least four times larger than \SI{6}{ms}, which is the time difference in the line scanning confocal between acquiring the upper and lower vesicle boundaries. The fit was performed with a fixed membrane bending rigidity $\kappa$ of \SI{21}{k_BT}. This yielded a membrane tension $\sigma$ of \SI{94+-7}{nN/m}. The analysis was done according to P\'ecr\'eaux et al., Eur. Phys. J. E \textbf{13} (2004), 277.}
\end{figure}

\begin{figure}[H]
	\centering{\includegraphics{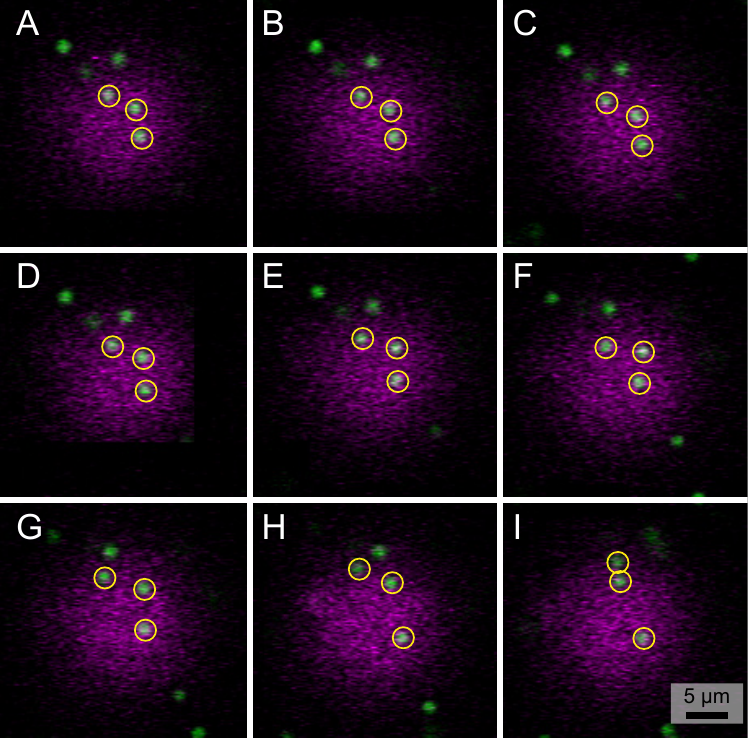}}
	\caption{Time lapse of the Brownian motion of membrane-attached particles on top of a vesicle. Three tracked particles are marked with yellow circles. Frames 0, 1, 2, 4, 8, 16, 32, 64, and 128 are displayed in \textit{(A)} to \textit{(I)}, respectively. The imaging frame rate was \SI{57}{Hz}.}
\end{figure}

\begin{figure}[H]
	\centering{\includegraphics[width=0.8\textwidth]{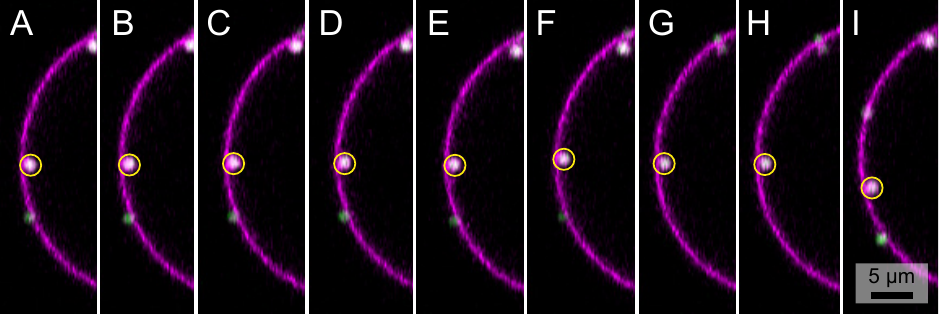}}
	\caption{Time lapse of the Brownian motion of a membrane-wrapped particle in the equatorial plane of a vesicle. The particle is marked with a yellow circle. Frames 0, 1, 2, 4, 8, 16, 71, 128, and 256 are displayed in \textit{(A)} to \textit{(I)}, respectively. The imaging frame rate was \SI{57}{Hz}.}
\end{figure}